\documentclass[12pt, preprint]{aastex}

\slugcomment{Accepted for Publication in {\it The Astrophysical Journal}}

\shorttitle{$\gamma\gamma$ Absorption by BLR Radiation Fields}
\shortauthors{M. B\"ottcher and P. Els}

\begin{document}

\title{Gamma-Gamma Absorption in the Broad Line Region Radiation Fields 
of Gamma-Ray Blazars}

\author{Markus B\"ottcher\altaffilmark{1,2}, and Paul Els\altaffilmark{1}}

\altaffiltext{1}{Centre for Space Research, North-West University, Potchefstroom,
2531, South Africa}

\altaffiltext{2}{Astrophysical Institute, Department of Physics and Astronomy, \\
Ohio University, Athens, OH 45701, USA}

\begin{abstract}
The expected level of $\gamma\gamma$ absorption in the Broad Line Region (BLR) 
radiation field of $\gamma$-ray loud Flat Spectrum Radio Quasars (FSRQs)
is evaluated as a function of the location of the $\gamma$-ray emission region.
This is done self-consistently with parameters inferred from the shape of the 
spectral energy distribution (SED) in a single-zone leptonic EC-BLR model scenario.
We take into account all geometrical effects both in the calculation of the 
$\gamma\gamma$ opacity and the normalization of the BLR radiation energy density. As 
specific examples, we study the FSRQs 3C279 and PKS 1510-089, keeping the BLR 
radiation energy density at the location of the emission region fixed at the 
values inferred from the SED. We confirm previous 
findings that the optical depth due to $\gamma\gamma$ absorption in the BLR radiation 
field exceeds unity for both 3C279 and PKS 1510-089 for locations of the $\gamma$-ray
emission region inside the inner boundary of the BLR. It decreases monotonically, 
with distance from the central engine and drops below unity for locations within 
the BLR. For locations outside the BLR, the BLR radiation energy density required 
for the production of GeV $\gamma$-rays rapidly increases beyond observational 
constraints, thus making the EC-BLR mechanism implausible. Therefore, in order to 
avoid significant $\gamma\gamma$ absorption by the BLR radiation field, the $\gamma$-ray 
emission region must therefore be located near the outer bounary of the BLR.
\end{abstract}

\section{Introduction}

Blazars are a class of radio-loud, jet-dominated active galactic nuclei whose
jets are oriented at a small angle with respect to our line of sight. Their
broadband emission is characterized by two broad non-thermal radiation components,
from radio to UV/X-rays, and from X-rays to $\gamma$-rays, respectively. The low
energy emission is generally understood to be due to synchrotron radiation by
relativistic electrons in a localized emission region in the jet. In leptonic
models for the high-energy emission of blazars \citep[see, e.g.,][for a discussion
of the alternative, hadronic models]{Boettcher13}, the $\gamma$-ray emission 
is due to Compton upscattering of soft target photon fields by the same
ultrarelativistic electrons in the jet. In the case of low-frequency-peaked
blazars (with synchrotron peak frequencies typically below $\sim 10^{14}$~Hz),
such as Flat Spectrum Radio Quasars (FSRQ), which show strong optical -- UV 
emission lines from a Broad Line Region (BLR), it is often argued that the
target photons for $\gamma$-ray production are the external (to the jet)
photons from the BLR \citep[e.g.,][]{Madejski99}. This would naturally suggest 
that the $\gamma$-ray emission region is located inside the BLR, in order to 
experience a sufficiently high radiation energy density of this target photon 
field. 

This picture, however, seems to be challenged by the detection of several FSRQs 
\citep[including 3C279, PKS 1510-089: ][]{Albert08,Abramowski13} as sources of 
very-high-energy (VHE, $E > 100$~GeV) $\gamma$-rays: VHE $\gamma$-rays produced
in the intense BLR radiation fields of these FSRQs are expected to be subject to
$\gamma\gamma$ absorption \citep[e.g.][]{Donea03,Reimer07,Liu08,Sitarek08,Boettcher09}.
This has repeatedly been considered as evidence that the $\gamma$-ray emission region 
must be located near the outer edge of the BLR \citep[e.g.,][]{Tavecchio11}, in order 
to avoid excess $\gamma\gamma$ absorption by the BLR radiation field, or that exotic 
processes, such as photon to Axion-Like Particle conversion, may act to suppress the 
impact of $\gamma\gamma$ absorption \citep[e.g.,][]{Tavecchio12}. 

The above referenced works on the $\gamma\gamma$ opacity due to the BLR radiation field,
however, used generic parameters for the respective FSRQs, independent of parameters
and emission scenarios actually required for the production of the observed $\gamma$-ray
emission in those blazars. In this paper, we consider two VHE $\gamma$-ray detected FSRQs, 
namely 3C279 and PKS 1510-089. We start out with constraints on the BLR luminosity 
and energy density from direct observations, 
under the assumption that the MeV -- GeV $\gamma$-ray emission is the result of
Compton upscattering of the BLR radiation field (EC-BLR) by the same ultrarelativistic 
electrons responsible for the IR -- optical -- UV synchrotron emission.
Within the observational constraints, we then self-consistently investigate 
the dependence of the $\gamma\gamma$ opacity due to the BLR radiation field on the location 
of the $\gamma$-ray emission region. This is done by re-normalizing the local emissivity 
in the BLR (within the observational constraints) for any given location of the $\gamma$-ray 
emission region to result in the required energy density experienced by the emission region, 
which is kept fixed in the process. 

In Section \ref{Model}, we describe the general model setup and methodology of our
calculations. Section \ref{Results} presents the results, specifically for 3C279 
(Section \ref{3C279}) and PKS 1510-089 (Section \ref{PKS1510}). Section \ref{Summary}
contains a brief summary and a discussion of our results.

\section{\label{Model}Model Setup}

\begin{figure}[ht]
\centering
\includegraphics[width=12cm]{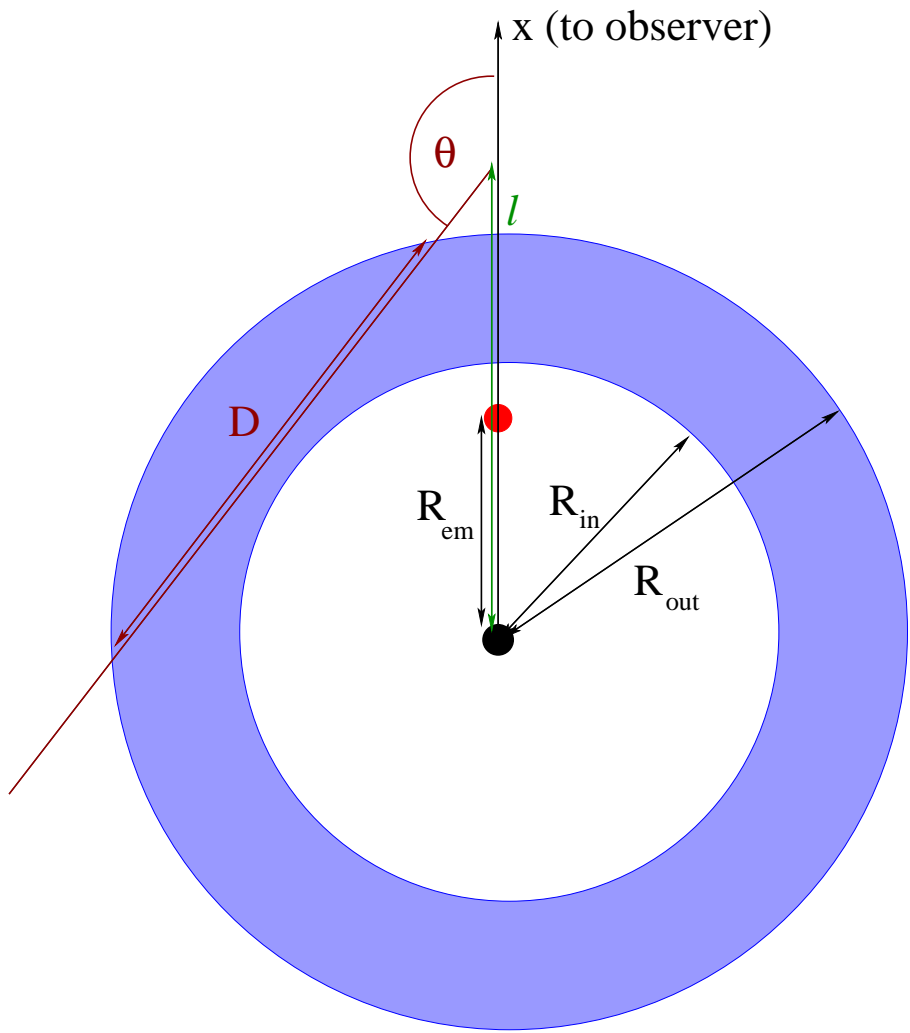}
\caption{\label{Geometry}Illustration of the model geometry used for the 
BLR $\gamma\gamma$ opacity calculation. }
\end{figure}

Our considerations are based on the frequently used model assumption that the 
$\gamma$-ray emission from FSRQ-type blazars is the result of the EC-BLR mechanism 
\citep[e.g.,][]{Ghisellini10,Boettcher13}. We represent the BLR as a spherical, 
homogeneous shell locally emitting with an emissivity $j_{\epsilon}^0$ within an 
inner ($R_{\rm in}$) and outer ($R_{\rm out}$) boundary of the BLR. The geometry 
of our calculations is illustrated in Figure \ref{Geometry}. 

Under the single-zone leptonic model assumptions with the EC-BLR mechanism producing
the MeV -- GeV $\gamma$-ray emission, the energy density of the BLR can be uniquely
determined solely based on the peak frequencies and $\nu F_{\nu}$ peak fluxes of the
synchrotron and EC $\gamma$-ray components of the SED. For this purpose, we make the
simplifying assumption that the Doppler factor $\delta = \left( \Gamma [1 - \beta_{\Gamma}
\cos\theta_{\rm obs}] \right)^{-1}$ is equal to the bulk Lorentz factor $\Gamma$ 
(corresponding to a normalized velocity $\beta_{\Gamma} = \sqrt{1 - 1/\Gamma^2}$) of
the flow, which is true to within a factor of $\lesssim 2$ for blazars, in which we 
are viewing the jet at a small observing angle $\theta_{\rm obs} \lesssim 1/\Gamma$. 
We furthermore assume that the $\gamma$-ray peak in the SED is dominated by Compton 
upscattering of Ly$\alpha$ photons from the BLR in the Thomson regime. This latter 
assumption is valid for FSRQ-type blazars in which the $\gamma$-ray peak typically
occurs at $E < 1$~GeV (and which we are considering in this paper), but may not hold 
for blazars of the intermediate- or high-frequency peaked classes. In the following,
photon energies are expressed as dimensionless values $\epsilon = h \nu / (m_e c^2)$.

The synchrotron peak frequency in the blazar SED is then given by $\nu_{\rm sy} \approx 
\nu_0 \, B_{\rm G} \, \gamma_p^2 \, \Gamma / (1 + z)$, where $\nu_0 \approx 4 \times 10^6$~Hz, 
$B_{\rm G}$ is the magnetic field in the emission region in units of Gauss, and $\gamma_p$ 
is the Lorentz factor of electrons radiating at the peak of the SED (i.e., the peak of the 
electron energy spectrum in a $\gamma^2 \, n(\gamma)$ representation). The EC-BLR peak 
frequency is located at $\epsilon_{\rm EC} \approx \epsilon_{Ly\alpha} \, \gamma_p^2 \, 
\Gamma^2 / (1 + z)$, where $\epsilon_{Ly\alpha} \approx 2 \times 10^{-5}$. These two 
observables can be used to constrain the magnetic field:

\begin{equation}
B_{\rm G} = {\nu_{\rm sy} \over \nu_0} \, {\epsilon_{Ly\alpha} \over \epsilon_{\rm EC}}
\, \Gamma 
\label{Bestimate}
\end{equation}

Denoting $f_{\rm sy / EC}$ as the peak $\nu F_{\nu}$ flux values of the synchrotron and
EC-BLR components, respectively, the ratio of EC-BLR to synchrotron peak $\nu F_{\nu}$ 
fluxes may then be used to constrain the BLR radiation energy density, since 

\begin{equation}
{f_{\rm EC} \over f_{\rm sy}} \approx {8 \, \pi \, u_{\rm BLR} \, \Gamma^2 \over B^2}
\label{fratio}
\end{equation}
which finally yields

\begin{equation}
u_{\rm BLR} \approx {1 \over 8 \pi} \, {f_{\rm EC} \over f_{\rm sy}} \, 
\left( {\nu_{\rm sy} \over \nu_0} \, {\epsilon_{Ly\alpha} \over \epsilon_{\rm EC}} \right)^2
\; {\rm erg} \; {\rm cm}^{-3}
\label{ublr}
\end{equation}

Notably, the dependence on the uncertain bulk Lorentz (and Doppler) factor cancels out in this 
derivation, so that Equation \ref{ublr} provides a rather robust estimate of $u_{\rm BLR}$ in
the framework of a single-zone leptonic EC-BLR interpretation of the blazar SED. 

It has been shown \citep{TG08,Boettcher13} that the $\gamma$-ray spectrum
resulting from Compton upscattering of a thermal blackbody at a temperature of $T_{\rm BLR}
= 2 \times 10^4$~K is an excellent approximation to the spectrum calculated with a detailed, 
line-dominated BLR spectrum. 
However, $\gamma\gamma$ absorption features are known to be much more sensitive to the
exact shape of the target photon spectrum. Therefore, for our evaluation of the $\gamma\gamma$
opacity in the BLR radiation field, we use a detailed, line-dominated BLR spectrum 
including the 21 strongest optical and UV emission lines with wavelengths and 
relative fluxes as listed in \cite{Francis91}.

Based on the value of $u_{\rm BLR}$ estimated through Equation \ref{ublr}
and observational 
constraints on the BLR luminosity $L_{\rm BLR}$, we first estimate the approximate location of 
the BLR, $R_{\rm BLR}$ through 

\begin{equation}
R_{\rm BLR} = \sqrt{ L_{\rm BLR} \over 4 \pi \, u_{\rm BLR} \, c}
\label{RBLR}
\end{equation}
$L_{\rm BLR}$ is either directly measured or estimated to be a fraction 
($f \sim 0.01$ -- 0.1) of the accretion-disk luminosity. The boundaries 
of the BLR are then chosen as $R_1 = 0.9 \, R_{\rm BLR}$ and $R_2 = 1.1 \, R_{\rm BLR}$.
We have done calculations with different widths of the BLR and verified that the choice 
of these boundary radii has a negligible influence on our final results. 

For any given location of the emission region at a distance $R_{\rm em}$ from the central 
supermassive black hole of the AGN, the emissivity $j_{\epsilon}$ at any point within the 
BLR is then fixed through the normalization to the required energy density $u_{\rm BLR}$ 
as resulting from a proper angular integration, assuming azimuthal symmetry around the $x$ 
axis:

$$
u_{\rm BLR} = \int\limits_0^{\infty} d\epsilon \, \int\limits_0^{\infty} dr \, 
2 \pi \int\limits_{-1}^{1} r^2 \, d\mu \; {j_{\epsilon} (\overrightarrow{r}) 
\over 4 \pi \, r^2 \, c} $$
\begin{equation}
= {1 \over 2 c} \int\limits_0^{\infty} d\epsilon \, j_{\epsilon}^0 \,
\int\limits_{-1}^1 d\mu \, D(\mu)
\label{uBLRnorm}
\end{equation}
where $j_{\epsilon} (\overrightarrow{r})$ is a Heaviside function equal to $j_{\epsilon}^0$
for locations $\overrightarrow{r}$ inside the BLR (i.e., between $R_{\rm in}$ and $R_{\rm out}$), 
and 0 elsewhere, and $D (\mu)$ is the length of the light path through the BLR in any given 
direction $\mu = \cos\theta$ (see Figure \ref{Geometry}). Once the normalization $j_{\epsilon}^0$ 
of the BLR emissivity is known, the $\gamma\gamma$ opacity for $\gamma$-rays emitted at the 
location $R_{\rm em}$ along the $x$ axis is calculated as

\begin{equation}
\tau_{\gamma\gamma} (\epsilon_{\gamma}) = {1 \over 2 c} 
\int\limits_{R_{\rm em}}^{\infty} d l \, \int\limits_{-1}^1 d\mu \, \int\limits_0^{\infty} 
d\epsilon \, {j_{\epsilon}^0 \, D (\mu) \over \epsilon \, m_e c^2 } \, (1 - \mu_i) \, 
\sigma_{\gamma\gamma} (\epsilon_{\gamma}, \epsilon, \mu_i)
\label{taugg}
\end{equation}
where $\mu_i = - \mu$ is the cosine of the interaction angle between the $\gamma$-ray and
the BLR photon, and $\sigma_{\gamma\gamma}$ is the polarization-averaged $\gamma\gamma$ 
absorption cross section:

\begin{equation}
\sigma_{\gamma\gamma} (\epsilon_{\gamma}, \epsilon, \mu_i) = {3 \over 16} \, \sigma_T \,
(1 - \beta_{\rm cm}^2) \, \left( \left[ 3 - \beta_{\rm cm}^4 \right] \, \ln \left[ {1 + 
\beta_{\rm cm} \over 1 - \beta_{\rm cm}} \right] - 2 \beta_{\rm cm} \, \left[ 2 - \beta_{\rm cm}^2
\right] \right)
\label{sigmagg}
\end{equation}
\citep{Jauch76} where $\beta_{\rm cm} = \sqrt{ 1 - 2 / (\epsilon_{\gamma} \epsilon \, [1 - \mu_i])}$.

It is obvious that the re-normalization of the local emissivith $j_{\epsilon}^0$ depending on the
location of the $\gamma$-ray emission region (according to Equ. \ref{uBLRnorm}), implies that the 
inferred BLR luminosity,

\begin{equation}
L_{\rm BLR}^{\rm requ} = {4 \over 3} \pi \left( R_{\rm out}^3 - R_{\rm in}^3 \right) \, 
\int\limits_0^{\infty} j_{\epsilon}^0 \, d\epsilon
\label{LBLR}
\end{equation}
may deviate from the observationally determined value. In particular, $L_{\rm BLR}^{\rm requ}$
will increase rapidly for locations of the $\gamma$-ray emission region outside of $R_{\rm out}$
(in order to keep $u_{\rm BLR}$ constant). We consequently restrict our considerations to a 
range of $R_{\rm em}$ within which $L_{\rm BLR}^{\rm requ}$ is within plausible observational 
uncertainties of the reference value.

\begin{figure}[ht]
\vskip 0.5cm
\centering
\includegraphics[width=12cm]{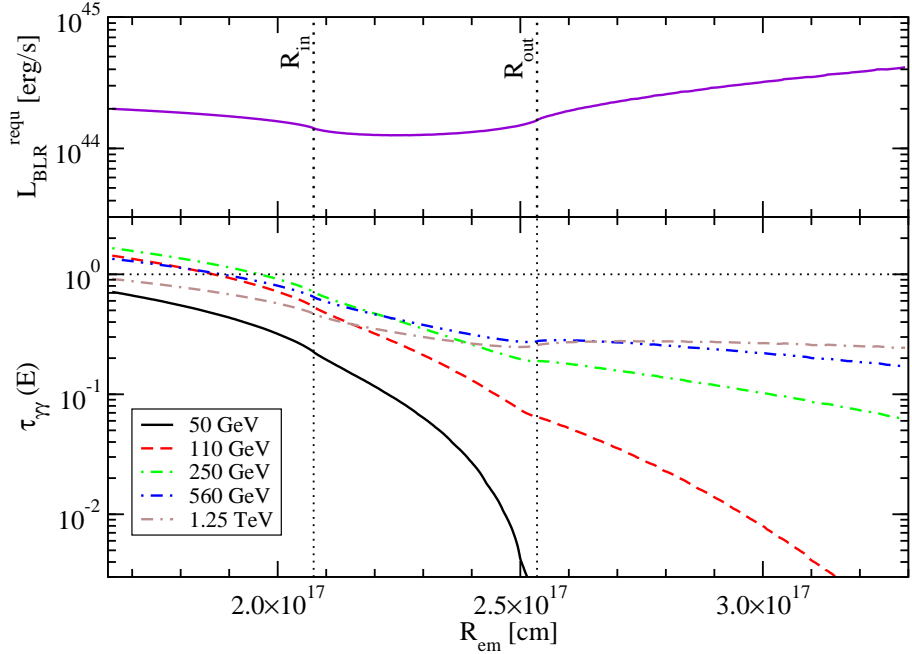}
\caption{\label{3C279fig} Results for 3C279. {\it Lower Panel:} $\gamma\gamma$ absorption optical 
depth as a function of location of the emission region, $R_{\rm em}$, for a fixed value of 
$u_{\rm BLR}$ as encountered by the emission region at the respective location (see text), for
several $\gamma$-ray photon energies. {\it Upper Panel:} Required luminosity of the BLR, according
to the re-normalization of the local BLR emissivity (Equation \ref{LBLR}). }
\end{figure}

\section{\label{Results}Results}

\subsection{\label{3C279}3C279}

The BLR luminosity of 3C279 was estimated by \cite{pian05} to be $L_{\rm BLR}^{\rm obs} = 
2 \times 10^{44}$~erg~s$^{-1}$. 
Representative SEDs of 3C279 \citep[e.g.,][]{Abdo10} show a synchrotron peak frequency
of $\nu_{\rm sy} \sim 10^{13}$~Hz and a $\gamma$-ray (EC-BLR) peak energy of $\epsilon_{\rm EC}
\sim 10^2$, while the $\gamma$-ray to synchrotron flux ratio is characteristically $f_{\rm EC}
/ f_{\rm sy} \sim 5$. This yields an estimate of the BLR radiation energy density of 
$u_{\rm BLR} = 1 \times 10^{-2}$~erg~cm$^{-3}$, implying an average radius of the BLR
(according to Equ. \ref{RBLR}) of $R_{\rm BLR} = 2.3 \times 10^{17}$~cm.

Figure \ref{3C279fig} illustrates the resulting $\gamma\gamma$ optical depth due to the 
BLR radiation field for various $\gamma$-ray photon energies (lower panel) and the required 
BLR luminosity (upper panel) as a function of the location of the $\gamma$-ray emission region. 
For most photons in the VHE $\gamma$-ray regime, the $\gamma\gamma$ opacity exceeds one for
locations far inside the inner boundary of the BLR, and gradually drops to values slightly
below one when approaching the BLR. 

It is well known \citep[e.g.][]{BD98} that, for a fixed emissivity (and, hence, luminosity)
of the BLR, the BLR photon 
energy density slowly increases when approaching the inner boundary of the BLR. Consequently, 
as we keep $u_{\rm BLR}$ fixed in our procedure, the inferred BLR luminosity has to decrease
as we consider locations of the emission region closer to $R_{\rm in}$, which adds to the
effect of a decreasing optical depth simply due to the decreasing path length of the $\gamma$-ray
photons through the BLR radiation field. The opacity continues to decrease as the emission region
is located inside the BLR. Notably, the decrease of $\tau_{\gamma\gamma}$ for locations outside 
the BLR is very shallow, at least for photons at $E \gg 100$~GeV, because the fixed value of 
$u_{\rm BLR}$ requires a rapidly increasing local emissivity $j_{\epsilon}^0$ (and, thus,
BLR luminosity). For this reason, we quickly reach values of $L_{\rm BLR}^{\rm requ.} \sim 2 \,
L_{\rm BLR}^{\rm obs}$ which we consider excessive compared to the observationally determined 
value. Thus, if the $\gamma$-ray emission region is located beyond the distance range 
considered in Figure \ref{3C279fig}, the GeV $\gamma$-ray emission can no longer be
produced by EC scattering of BLR photons with plausible parameter choices, and would,
instead, have to be produced by a different mechanism, such as EC scattering of IR
photons from a dusty torus.

\begin{figure}[ht]
\centering
\vskip 0.4cm
\includegraphics[width=12cm]{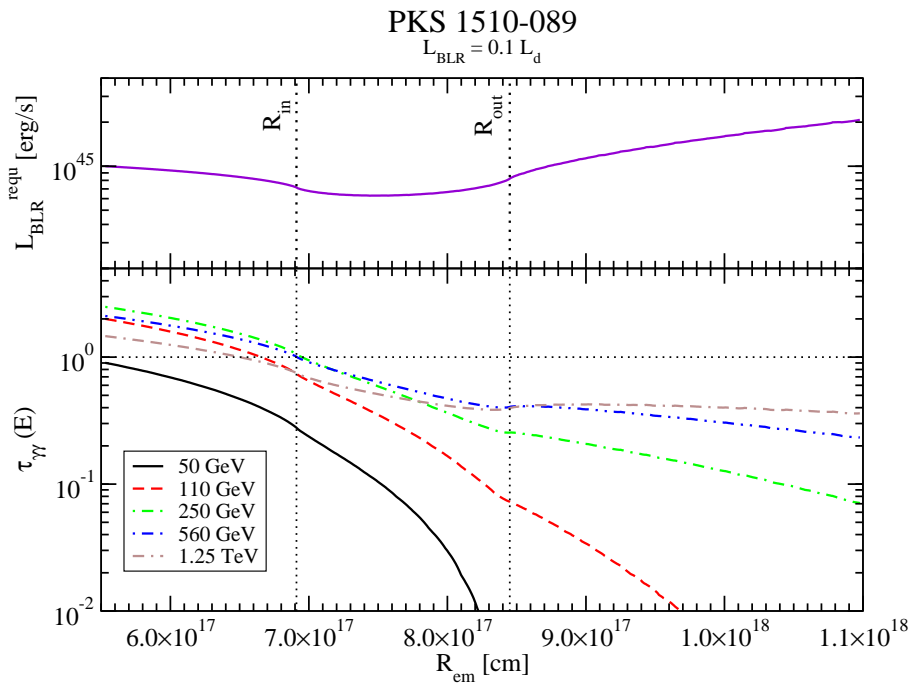}
\caption{\label{PKS1510_01} Results for PKS 1510-089, assuming $L_{\rm BLR} = 0.1 \, L_d$. 
Panels and symbols as in Figure \ref{3C279fig}. }
\end{figure}

\subsection{\label{PKS1510}PKS 1510-089}

In the case of PKS 1510-089, to our knowledge, no value of the total luminosity of the BLR
has been published. We therefore parameterize the luminosity of the BLR as a fraction $f =
0.1 f_{-1}$ of the accretion disk, $L_{\rm BLR} = f \, L_d$. The accretion disk luminosity
was determined by \cite{pucella08} to be $L_d = 1.0 \times 10^{46}$~erg~s$^{-1}$. 
Characteristic SEDs of PKS 1510-089 \citep[e.g.,][]{Abdo10} indicate $\nu_{\rm sy}
\sim 3 \times 10^{12}$~Hz, $\epsilon_{\rm EC} \sim 10^2$, and $f_{\rm EC} / f_{\rm sy}
\sim 20$, for which Equation \ref{ublr} yields $u_{\rm BLR} = 4.5 \times 10^{-3}$~erg~cm$^{-3}$, 
yielding a BLR radius of $R_{\rm BLR} = 7.7 \times 10^{17} \, f_{-1}^{1/2}$~cm.

The results for a fiducial value of $f = 0.1$ (i.e., BLR luminosity = 10~\% of the accretion
disk luminosity) are illustrated in Figure \ref{PKS1510_01}. The general trends are the same
as found for 3C279,
with slightly larger values of $\tau_{\gamma\gamma}$ due to the larger BLR luminosity
(assuming $f = 0.1$) and larger BLR size. Still, the same conclusion holds: If the GeV
$\gamma$-rays are produced by the EC-BLR mechanism, the $\gamma$-ray emission region must
be located near the outer boundary of the BLR, whereas for locations far beyond the outer
boundary, the EC-BLR mechanism becomes implausible for the production of the observed GeV
$\gamma$-ray flux.

\begin{figure}[ht]
\centering
\vskip 0.4cm
\includegraphics[width=12cm]{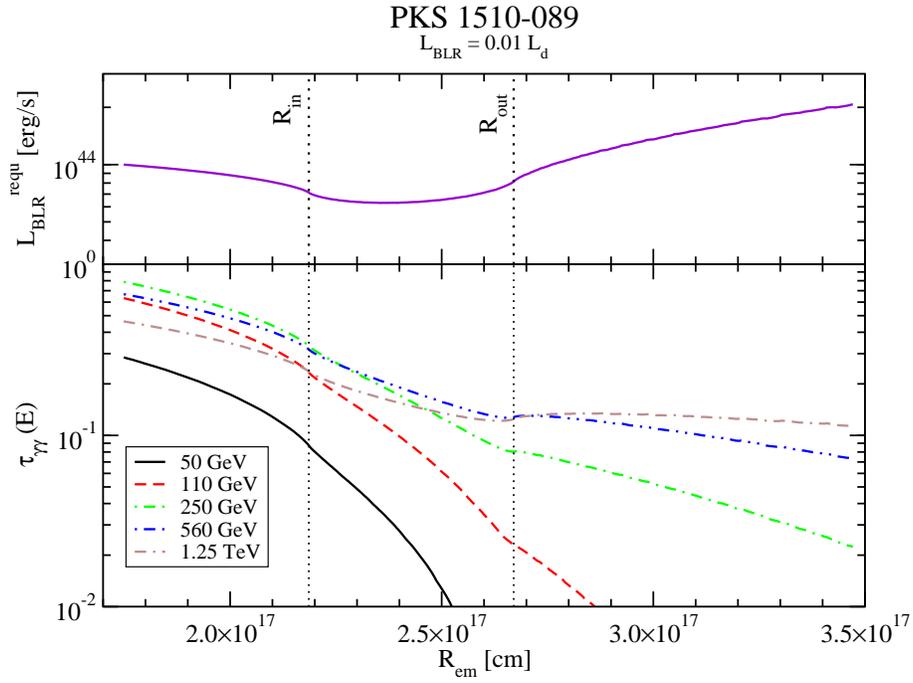}
\caption{\label{PKS1510_001} Results for PKS 1510-089, assuming $L_{\rm BLR} = 0.01 \, L_d$. 
Panels and symbols as in Figure \ref{3C279fig}. }
\end{figure}

Figure \ref{PKS1510_001} illustrates that this general result is is only weakly dependent
on the value of $f$, with $\gamma\gamma$ opacities being smaller for smaller values of $f$
(i.e., smaller values of $L_{\rm BLR}$, but keeping $u_{\rm BLR}$ fixed). This is expected 
as a smaller value of $L_{\rm BLR}$ implies a smaller size of the BLR and, thus, a smaller
effective path length of $\gamma$-ray photons through the BLR radiation field. Consequently,
an approximate scaling $\tau_{\gamma\gamma} \propto f^{1/2}$ holds.

\section{\label{Summary}Summary and Discussion}

We have re-evaluated the $\gamma\gamma$ opacity for VHE $\gamma$-rays in the BLR radiation 
fields of VHE-detected FSRQ-type $\gamma$-ray blazars. Our method started from a fixed value
of the radiation energy density $u_{\rm BLR}$ and inferred average radius of the BLR, based
on the observationally constrained BLR luminosity. Keeping the value of $u_{\rm BLR}$ fixed,
we calculated $\tau_{\gamma\gamma}$ for a range of locations of the $\gamma$-ray emission region,
from inside the inner boundary to outside the outer boundary of the BLR. For the specific examples
of 3C279 and PKS 1510-089, we found that the resulting $\gamma\gamma$ opacities for VHE $\gamma$-ray
photons exceed unity for locations of the $\gamma$-ray emission region inside the inner boundary
of the BLR (in the case of PKS 1510-089, this is true for $L_{\rm BLR} \gtrsim 0.1 \, L_d$), 
in agreement with previous studies \citep[e.g.,][]{Liu08,Sitarek08,Boettcher09}.
We find that, under the assumption of the GeV $\gamma$-ray emission being produced by the
EC-BLR mechanism, the $\gamma\gamma$ opacity gradually drops for locations of the 
$\gamma$-ray emission region approaching the BLR and within the boundary radii of the BLR,
reaching values far below unity when approaching the outer boundary. For locations outside 
the BLR, the BLR luminosity required to still be able to produce the observed GeV $\gamma$-ray 
flux through the EC-BLR mechanism, quickly exceeds observational constraints, thus requiring 
alternative $\gamma$-ray production mechanisms, such as EC scattering of IR photons from a 
dusty torus. Alternative radiation mechanisms / target photon fields are required in any case
for the production of VHE $\gamma$-rays, since Compton scattering of the optical/UV target 
photons from the BLR to $> 100$~GeV energies would occur in the Klein-Nishina regime, in
which this process is strongly suppressed.

In the case of PKS 1510-089, the uncertain BLR luminosity allows for configurations
of the VHE $\gamma$-ray emission region even within the inner boundary of the BLR if the 
BLR luminosity is $L_{\rm BLR} \lesssim 10^{-2} \, L_d$, i.e., in the case of a very small
covering factor of the BLR. 

The generic estimates of the BLR radiation energy density and inferred radius of the BLR
based on the SED characteristics and the assumption of $\gamma$-ray production dominated
by EC scattering of BLR photons, are in reasonable agreement with independent methods of
determining $R_{\rm BLR}$ (and, thus, $u_{\rm BLR}$). Specifically, \cite{pian05} estimated 
the size of the BLR of 3C279 to be $R_{\rm BLR} \sim 9 \times 10^{16}$~cm. \cite{Bentz09} 
provided a general scaling of the size of the BLR with the continuum luminosity of the 
accretion disk, $L_d = 10^{45} \, L_{d, 45}$~erg~s$^{-1}$, of $R_{\rm BLR} \sim 3 \times 
10^{17} \, L_{d, 45}^{1/2}$~cm, where the continuum lumonisity $\lambda L_{\lambda}$ at $\lambda =
5100$~\AA\ is used as a proxy for the disk luminosity. This implies a universal value of 
$u_{\rm BLR} \sim 3 \times 10^{-2} \, f$~erg~cm$^{-3}$, in reasonable agreement with our
SED-based estimates. 

The $\gamma\gamma$ opacity constraints derived here can, of course, be circumvented if
(a) the GeV $\gamma$-ray emission is not produced by the EC-BLR mechanism, or (b) the
GeV and TeV $\gamma$-ray emissions are not produced co-spatially. In case (a) the energy 
density of the BLR radiation field at the location of the $\gamma$-ray emission region 
can be arbitrarily small, i.e., the $\gamma$-rays can be produced at distances far
beyond the BLR. Evidence for $\gamma$-ray production at distances of tens of pc from
the central engine has been found in a few cases, based on correlated $\gamma$-ray and
mm-wave radio variability \citep[e.g.,][]{Agudo11}. In this case, GeV $\gamma$-rays can
still be produced in a leptonic single-zone EC scenario by Compton scattering external 
infrared radiation from a dusty torus. However, it is often found that, in order to
provide a satisfactory representation of the SEDs of FSRQ-type blazars, both the BLR
and the torus-IR radiation fields are required as targets for $\gamma$-ray production
\citep[e.g.,][]{Finke10}. In case (b), one would need to resort to multi-zone models,
in which the GeV emission could be produced within the BLR at sub-pc distances, but 
the VHE $\gamma$-rays are produced at distances of at least several parsecs. In such
a scenario, one would not expect a strong correlation between the variability patterns
at GeV and VHE $\gamma$-rays. This appears to be in conflict with the correlated 
GeV ({\it Fermi}-LAT) and VHE variability of PKS 1510-089 \citep{Abramowski13}
and PKS 1222+21 \citep{Aleksic11}, while the VHE $\gamma$-ray detections of 3C279 
by MAGIC \citep{Albert08} occurred before the launch of {\it Fermi}, so no statements 
concerning correlated GeV and VHE $\gamma$-ray variability can be made in this case.

\acknowledgements{We thank the anonymous referee for helpful comments which significantly
improved the paper. MB acknowledges support through the South African Research Chair Initiative 
(SARChI) of the Department of Science and Technology and the National Research Foundation\footnote{Any
opinion, finding and conclusion or recommendation expressed in this material is that of the authors,
and the NRF does not accept any liability in this regard.} of South Africa under NRF SARChI Chair 
grant no. 64789. }

\end{document}